 \documentclass[12pt,preprint]{aastex}

\begin{document}
\title{A Closer Look at the Soft X-ray Transient X1608$-$52: Long-term Optical/X-ray Observations} 
\author{Stefanie Wachter}  
\affil{Cerro Tololo Inter-American Observatory, National Optical Astronomy 
Observatory\altaffilmark{1}, Casilla 603, La Serena, Chile}
\altaffiltext{1}{Operated by the Association of Universities for Research in 
Astronomy, Inc., under cooperative agreement with the National Science 
Foundation.}
\email{swachter@noao.edu}

\author{D. W. Hoard}
\affil{Cerro Tololo Inter-American Observatory, National Optical Astronomy
Observatory\altaffilmark{1}, Casilla 603, La Serena, Chile}
\email{dhoard@noao.edu}

\author{Charles D. Bailyn}
\affil{Yale University, Dept. of Astronomy, PO Box 208101, 
New Haven CT 06520-8101}
\email{bailyn@astro.yale.edu}

\author{St\'ephane Corbel}
\affil{Universit\'e Paris VII and Service d'Astrophysique, CEA Saclay, 
F-91191 Gif sur Yvette, France}
\email{corbel@discovery.saclay.cea.fr}

\author{Philip Kaaret}
\affil{Harvard-Smithsonian Center for Astrophysics, 60 Garden Street, 
Cambridge, MA 02138}
\email{pkaaret@cfa.harvard.edu}

\begin{abstract}
We report long-term simultaneous optical and (RXTE) X-ray observations of the 
soft X-ray transient and low mass X-ray binary X1608$-$52 
spanning from 1999 to 2001. In addition to the usual X-ray outburst and 
quiescent states, X1608$-$52 also exhibits an extended low intensity state
during which the optical counterpart, QX~Nor, is found to be about two 
magnitudes brighter than during quiescence. We detect optical photometric 
variability on a possible period of 0.5370~days with a 
semi-amplitude of $\sim 0.27$
mag in the $I$ band. The modulation could be orbital but is also 
consistent with a 
scenario invoking a superhump with decreasing period.
We discuss the possibilities to distinguish between 
the orbital and superhump period cases. Observations of QX~Nor during 
quiescence indicate an F to G type main sequence secondary while
theoretical considerations argue for an evolved mass donor. Only an evolved 
mass donor would satisfy the condition for the occurrence of superhumps.

\end{abstract}


\section{Introduction}

Soft X-ray transients (SXTs) are a subclass of the low mass X-ray binaries (LMXBs),
mass-transfer systems where matter donated by a late-type star is
accreted onto either a neutron star (NS) or a black hole. SXTs
are characterized by prominent outbursts separated by long periods 
of quiescence. The physical mechanism underlying the outbursts remains 
under debate, although the favored scenario at present is some variation 
of the disk instability model (see the recent review by Lasota 2001). 
During an outburst, the luminosity of SXTs
increases by several orders of magnitude at all wavelengths.
While the optical light is dominated by the accretion disk and the heated 
face of the companion during
the outburst as in the persistent LMXBs, the quiescent state of
SXTs offers the rare opportunity to study the  
mass donor ``uncontaminated'' by other sources of emission.
Observations of the mass donor in quiescence can provide the
orbital period of the system through ellipsoidal variations,
the spectral type and radial velocity curve of the donor star, and
also constrain the mass of the compact object.
The majority of SXTs appear to contain black hole primaries and this group
provides some of the best black hole candidates with the most stringent 
mass constraints.

X1608$-$52 is a member of the rarer group
of SXTs with neutron star primaries (based on the detection of type I X-ray
bursts). X-ray outbursts from this system have been recorded since the 
early 1970's. While some of the data suggested a recurrence 
pattern of $\sim 600$ days (Murakami et al.\ 1980), Lochner \& Roussel-Dupre
(1994) reported a more complex outburst history. 
Hasinger \& van der Klis (1989) classified X1608$-$52 as an Atoll source. 
Kilohertz quasi-periodic oscillations (QPOs) were discovered during the 
decay of the 1996 outburst (van Paradijs et al.\ 1996, Berger et al.\ 1996),
twin QPO peaks with varying separation during the 1998 outburst 
(Mendez et al.\ 1998), and finally possible sideband oscillations 
(Jonker, Mendez \& van der Klis 2000).  

The optical counterpart of X1608$-$52, QX~Normae, was discovered by Grindlay \&
Liller (1978) after an outburst in 1977. 
Wachter (1997) detected QX~Nor in the IR and
optical a few months after its outburst in 1996.
If X1608$-$52 was to adhere to the recurrence pattern of its last two 
outbursts (February 1996 and 1998), the next outburst was expected to 
occur early in 2000. We report here on long-term optical and X-ray observations
spanning from 1999 to 2001. 

\section{Observations}

\subsection{Optical}

We performed CCD photometry of QX~Nor with various CTIO
telescopes between 1999 and 2001. A log of all observations can be found in 
Table~\ref{tab1}. The YALO 1m telescope is operated exclusively in queue
mode by a consortium consisting
of Yale, AURA, University of Lisbon, and Ohio State University. 
The SSTO (Synoptic/Service/Target--of--Opportunity)
Program at the CTIO 0.9m telescope provided queue scheduled observations
about once per week during Feb--Jul 2000.
Most observations of X1608$-$52 were carried out in the $I_C$ band 
with exposure times ranging from 600--900~s depending on the brightness
of the target and/or observing conditions. 
The data were overscan-corrected, bias-corrected, and flat-fielded in 
the standard 
manner using IRAF. Photometry was performed by point spread function fitting 
with DAOPHOT~II (Stetson 1993). The instrumental magnitudes were transformed to 
the standard system by comparison with previously calibrated local 
standards (Wachter 1997). The errors quoted here are the 
intrinsic 1$\sigma$ errors of this relative
photometry. The 1$\sigma$ error at the different magnitude levels was 
derived from the rms scatter in the
light curves of several comparison stars of appropriate brightness. 

\subsection{X-rays}

The bulk of the X-ray observations utilized in this paper are the 
publicly available data from the All Sky Monitor
\footnote{provided by the ASM/RXTE teams at MIT and at the RXTE SOF
and GOF at NASA's GSFC.} on board the  the Rossi X-ray Timing Explorer
(Bradt, Rothschild \& Swank 1993).
We also present data  obtained with the RXTE proportional counter array (PCA).
The PCA consists of five proportional counter units (PCUs) and is 
currently being operated with reduced duty cycles for three out of the 
five PCU detectors.
X1608$-$52 was observed (with the PCA) 34 times from 2000 March 6 to 
2000 May 24
and 14 times from 2001 January 2 to 2001 January 28. We also used 
the target-of-opportunity public domain data of X1608$-$52 from 2000 
January 6 to 2000 January 16 (4 observations). 

The observations were performed in various configurations; the X-ray data
presented here were obtained using Standard 1 and Standard 2 modes.
The Standard 1 mode has a time resolution of 1/8 s in one energy band 
(2-60 keV) and the Standard 2 mode has a time resolution of 16 s and 
129 channels in the 2-60 keV energy band. The Standard 1 data were used
to exclude time intervals from the light curve in which type I X-ray bursts
were present. We used the Standard 2 data to compute the intensity light
curve based on the various PCUs available, excluding PCU 0 after 2000 12 May 
(due to the loss of the propane veto layer in PCU 0). X-ray timing and spectral
analysis will be presented in a forthcoming paper.

We performed our data analysis using RXTE standard software FTOOLS 5.0 
and our own IDL routines. We only utilized data obtained when the source
was $> 10$ degrees above the Earth's limb. Intervals 
measured within 30 minutes of passages of the South Atlantic Anomaly 
or during times of high particle background (ELECTRON 0 or 2 greater 
than 0.1) were discarded. Data were only extracted from the top layer 
of the PCUs. Background subtraction was performed using the sky-VLE 
model for observations when X1608$-$52 was bright (net count rate 
greater than 40 c s$^{-1}$ PCU$^{-1}$) and the L7-240 model for the 
faintest intensity
observations.

\section{Low Intensity State versus True Quiescence}

As part of a monitoring program of SXTs in quiescence, we obtained 
several optical CCD frames of the X1608$-$52 field in May 1999. The ASM light curve 
indicated that the system was not active at the time and that $\sim13$
months had passed since its last outburst in 1998. Consequently, we 
expected the optical counterpart, QX Nor, to be undetectable as in 
previous observations during quiescence. To our surprise,
we observed the optical counterpart at the same brightness 
as in our observations only two months after the 1996 outburst
(Wachter 1997), at an average level of $I=18.6$, $R=20.1$! Wachter (1997)
had speculated that QX Nor was still bright two months after the outburst 
due to a delayed outburst decay in the optical compared to the X-rays. 
However, finding the counterpart
at similar brightness 13 months after an outburst clearly requires an 
alternative explanation. 

\begin{figure}
\epsscale{0.9}
\plotone{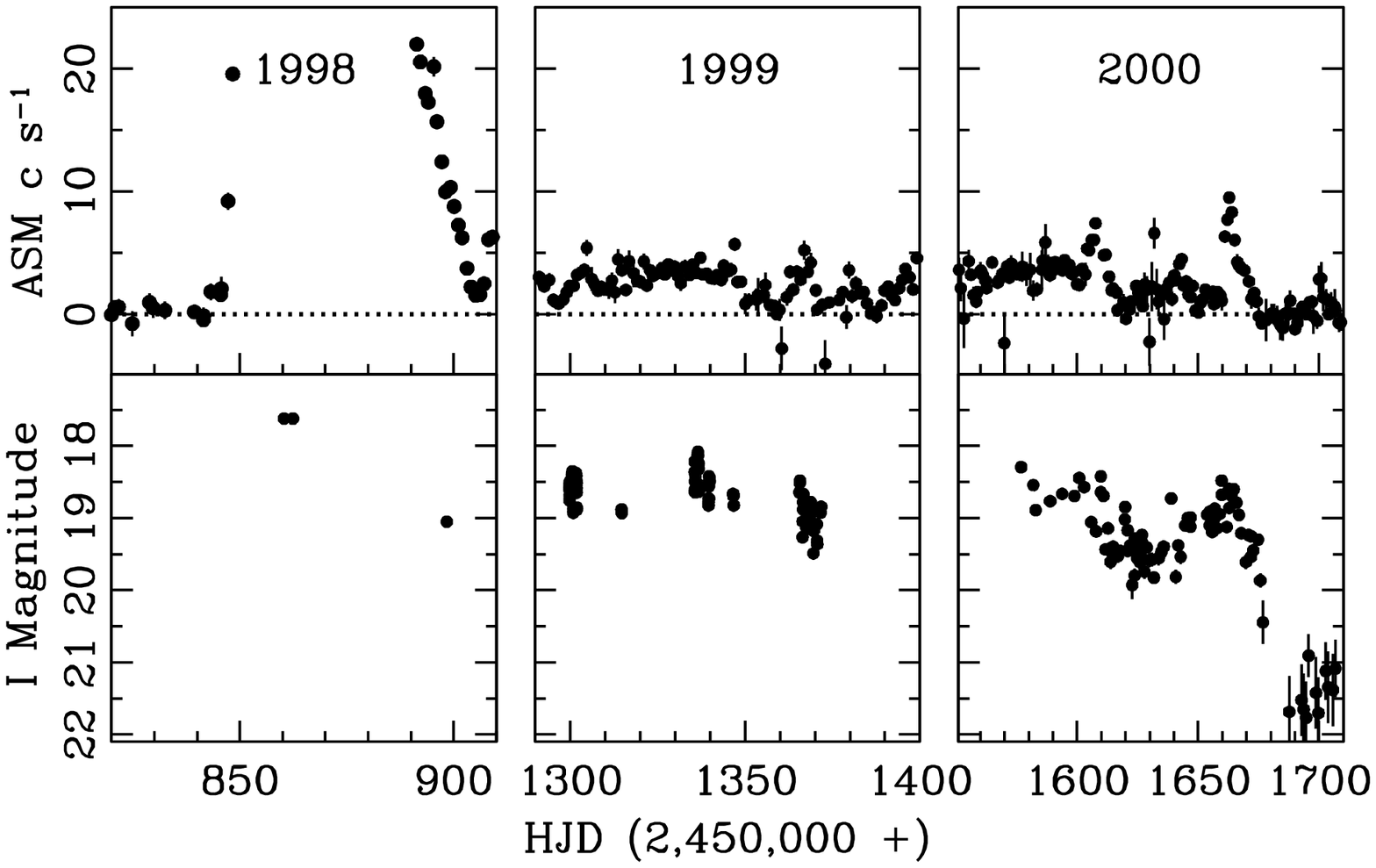}
\caption[wachter.f1.eps]{X-ray light curve ({\it top panels})
of X1608$-$52 from the RXTE ASM during time intervals with simultaneous 
optical coverage ({\it bottom panels}). The optical data during the 
1998 outburst are reproduced from Corbel \& Chapuis (2000). Notice 
the different source states: 
outburst ($\sim70$ ASM counts s$^{-1}$ at
peak -- not shown) with a corresponding optical brightness of $I=17.6$; 
LIS ($\sim3$ ASM counts s$^{-1}$) with $I=18.5-20.0$; TQ (after  
HJD 2,451,680 and immediately prior to the
1998 outburst, indicated by the dotted line) with the optical 
counterpart at $I=21 - 21.5$. 
\label{f-states}}
\end{figure}

After continued monitoring of QX Nor in June and July 1999 and the first half  
of 2000, it became clear that X1608$-$52 exhibits three distinct 
states: the outburst state, an extended low intensity state (LIS), and 
the quiescent state. We will refer to 
the quiescent state as true quiescence (TQ) in order to distinguish it from 
the LIS that was previously thought to also represent the quiescent 
state of the system. Figure~\ref{f-states} illustrates the various 
source states observed in the X-ray and optical. 
The only optical observations of QX Nor during 
an X-ray outburst have been reported by Corbel \& Chapuis (2000) and are 
reproduced in the figure. During the 1998 outburst 
($\sim70$ ASM counts s$^{-1}$ at 
peak), the optical counterpart reached $I=17.6$. During the 1999 and most 
of the 2000 observations, the source is in the LIS ($\sim 3$ ASM counts
s$^{-1}$) with a corresponding optical brightness of $I=18.5-20.0$. 
The variability apparent in the optical data in this state is partly due 
to the overall variation seen in the X-rays as well as possible orbital 
modulation (see Section 4).
Only after 2000 May 16 (HJD 2,451,680) and prior to the 
1998 outburst is the source at true quiescent levels. The ASM count rate 
hovers around zero and the optical counterpart drops to $I=21 - 21.5$. 
Our optical data during TQ constitute the first detection of the quiescent 
optical counterpart and will be discussed in more detail below. 

\begin{figure}
\epsscale{0.9}
\plotone{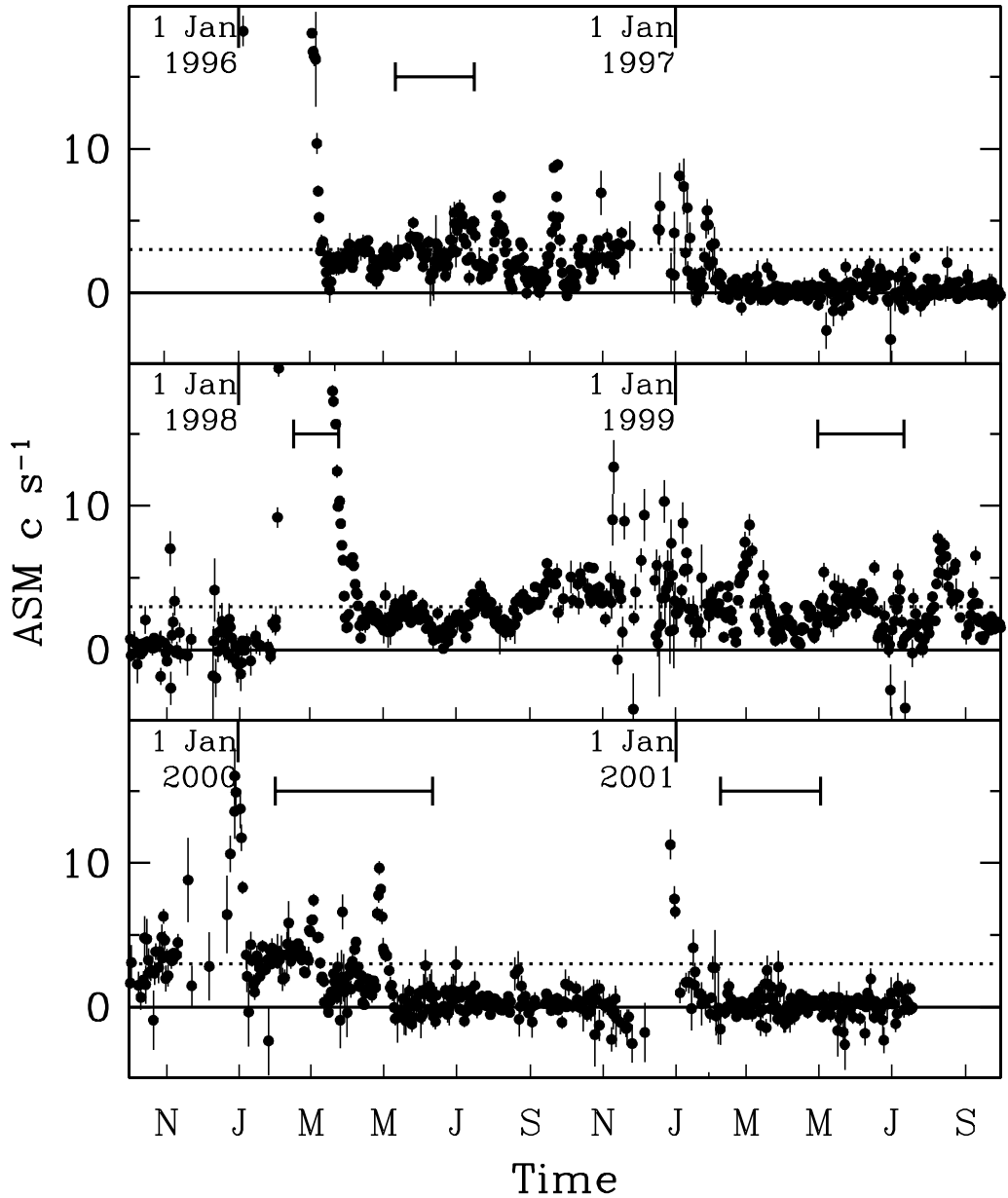}
\epsscale{1.00}
\caption[wachter.f2.eps]{The entire five year ASM light curve of
X1608$-$52. The times of optical coverage are marked by
horizontal bars. The peak X-ray count rate has been truncated to more 
clearly show the LIS and TQ. 
The X-ray count level of the LIS is indicated by the 
dotted line, the TQ by the solid
line. Each tickmark on the horizontal axis denotes a two month interval.
 \label{f-xray}}
\end{figure}

These different states of X1608$-$52 become even more apparent when 
considering the entire five year ASM light curve shown in Figure~\ref{f-xray}.  
After the outburst in 1996, X1608$-$52 remained in the LIS for $\approx11$ 
months, then spent $\approx12$ months in TQ, underwent a new outburst in 
1998 followed by almost two years in the LIS, and finally TQ for about 
the last year. In January of both 2000 and 2001, the ASM data seem to  
show the decaying tail of a new outburst of X1608$-$52. Unfortunately, 
all of those data points fall in the time interval each year (November 
to January) when the source is very close to the sun and the recorded 
data points are the most unreliable. These time intervals can be easily 
identified in the long-term X-ray light curve as times with  
missing coverage and/or increased scatter and measurement errors.
Even if all the data points can be trusted, the 2000 ``outburst'' would have 
been very short (there are several points that would correspond to 
the rise phase) compared to the previous two, and reached nowhere
near the previously recorded peak values. If real, it seems more reminiscent of 
one of the higher variability spikes evident in the LIS on various 
occasions, e.g.,
around October 1996, March 1999, and May 2000. In addition, in the context
of the outburst mechanism models, it might be more plausible for the system 
to replenish matter in the disk in TQ for some time before another 
major outburst can be sustained. The possible 2001 outburst is more 
difficult to evaluate. It would have started from TQ as expected, but also 
returned immediately 
into TQ as confirmed by the faintness of the optical counterpart. 
In contrast, X1608$-$52 remained in the LIS for extended periods
after both the 1996 and 1998 outbursts. Our 2001 PCA data do show 
count rates of 20--90 counts s$^{-1}$ PCU$^{-1}$ during January 2--19, 
supporting the authenticity of some X-ray activity in X1608$-$52. 
By the date of the first optical observation (2001 February 1), however, 
the system had returned to TQ. It remains unclear, whether the 2001 event 
constituted a major outburst of the system akin to those in 1996 and 1998.
We note that there are marked differences in the X-ray timing and 
spectral behavior 
between the outbursts in 2000 and 2001. For example, the low frequency noise 
in 2001 is much stronger than in 2000. This is true throughout 
the two outbursts. A detailed analysis of the X-ray data will be 
presented in a forthcoming paper.

\begin{figure}
\epsscale{0.9}
\plotone{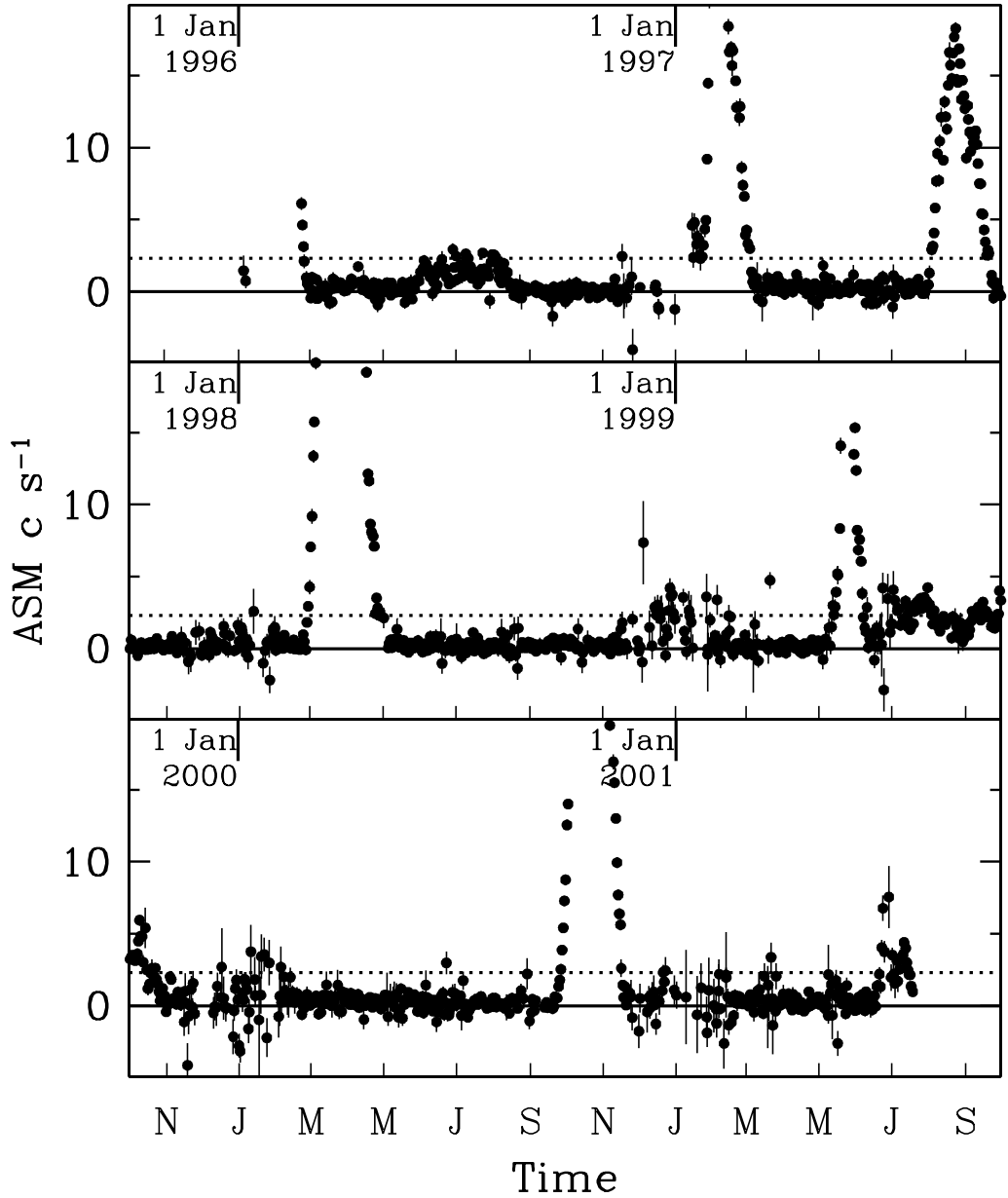}
\epsscale{1.00}
\caption[wachter.f3.eps]{The entire five year ASM light curve of
Aql X-1, equivalent to Figure~\ref{f-xray}. Note the elevated states
after the 1999 outburst and between outbursts in 1996, possibly 
corresponding to the LIS seen in X1608$-$52.
 \label{f-aqlx1}}
\end{figure}

Are the two non-outburst states, the LIS and TQ, unique features of X1608$-$52
or a universal characteristic of SXTs? The latter possibility might offer 
some additional clues to the physics underlying the outburst mechanism.
Aql X-1 is the only neutron star SXT other than X1608$-$52 that has undergone
multiple outbursts since the launch of RXTE and has sufficient
X-ray flux to distinguish between possible LIS and TQ states. The ASM lightcurve
from this source is shown in Figure~\ref{f-aqlx1}. On occasion, Aql X-1 
indeed seems to show the same two level states outside of outburst as observed
in X1608$-$52: a LIS is seen for several months after
the outburst in June 1999 as well as possibly between outbursts in 1996.
Jain et al.\ (2001) report optical observations of Aql X-1
across the 1999 outburst and LIS. The behavior of the optical counterpart 
is strikingly similar to that of X1608$-$52: the Jain et al.\ (2001) 
light curve shows a persistent optical brightness  
at a level $\approx 1$ magnitude lower than during the peak of the 
outburst, but several magnitudes brighter than quiescence, during the 
period July--November 1999, immediately following the large X-ray outburst. 
We can also compare the X-ray luminosities of the various states for 
both of these sources. Chen, Shrader, \& Livio (1997) list outburst 
peak / quiescent 
luminosities of about $(2-8)\times 10^{37}$ / $2\times 10^{33}$ ergs s$^{-1}$ 
for X1608$-$52 and $(1-5)\times 10^{37}$ / $6\times 10^{32}$ ergs s$^{-1}$ for 
Aql X-1. For X1608$-$52, our PCA data give $1.5\times 10^{35} - 4.6\times 
10^{36}$ ergs s$^{-1}$ 
for the LIS luminosity in the 2-10 keV band. Jain et al.\ (2001) arrive 
at $5\times 10^{35}$ ergs s$^{-1}$ for the LIS in Aql X-1 (the assumed 
distances for X1608$-$52 and Aql X-1 are 3.6 kpc and 2.5 kpc, respectively).  
Thus, the LIS is very similar in both sources. Interestingly,
most outbursts of Aql X-1 do not lead to an LIS, and there is
evidence for LISs unassociated with a true outburst state.  Most
recently, Bailyn et al.\ (2001) have reported an optical
brightening of the Aql X-1 counterpart starting in June 2001, 
during which the ASM flux never
exceeded 8 counts s$^{-1}$. This LIS seems likely to be analogous
to the modest X-ray activity observed in June 1996 and January 1999.

It is difficult to understand the LIS and TQ in the 
context of the pure disk instability model which describes the SXT phenomenon as 
a limit cycle between two discrete disk states. An outburst occurs when
a heating front in the disk is triggered and initiates the transition from a 
quiescent, cold, low viscosity disk to a hot, high viscosity state. 
Subsequent cooling fronts return the disk to the quiescent state (e.g. Lasota
2001).  
The model does not produce an extended intermediate state analogous to 
the observed LIS.
During the LIS, X1608$-$52 appears very much like a persistent LMXB, with a
relatively stable X-ray flux and photometrically varying counterpart.  
Instead of returning directly to the quiescent cold disk after an outburst, 
X1608$-$52 seems to 
retain a luminous accretion disk with a stable mass transfer rate sufficient
to sustain an elevated X-ray flux for an extended period of time. The 
presence of a luminous accretion disk during the LIS is supported by 
the observation that the brightness of the 
optical counterpart does not decrease as strongly from outburst state
to the LIS as from LIS to TQ.  
Figure~\ref{f-states} shows that the optical counterpart 
increases in brightness by about two magnitudes from TQ to LIS, but at  
most one magnitude between the LIS and the peak of the outburst 
(when the X-ray flux increases by a factor of more than 20). A similarly
small optical brightness difference between the LIS and full outburst level is 
seen in the 1999 Aql X-1 outburst. This implies 
that the main optically luminous components that are created in the system 
during the outburst are retained for the duration of the LIS.
The major elements likely to change during 
an SXT outburst are the size and luminosity of the accretion disk. 
Any additional brightening during outburst in the optical is probably 
due to an increase in X-ray 
irradiation of the disk and the mass donor and represents a smaller 
contribution to the overall brightness of the system 
than the transition from the cold quiescent disk to the hot disk during 
outburst and the LIS.   

The LIS is reminiscent of the standstills seen in the optical light curves of
Z Cam type cataclysmic variables (CVs). For Z Cam systems, 
the decline from outburst is interrupted 
and the systems remain at about 0.7 mag below the peak brightness 
(e.g. Warner 1995, p.\ 163). The 
duration of the standstills varies from ten days to years. After a standstill, 
the system returns to its usual quiescent state. These observed 
properties are interpreted in the framework of the disk instability model
where the mass transfer
rate from the secondary fluctuates about the critical rate that separates
stable systems from those that undergo dwarf nova outbursts. Buat-Menard, 
Hameury, 
\& Lasota (2001) show that heating of the outer disk by the accretion stream 
impact point also has to be included to successfully model observed Z Cam 
type lightcurves. Stellar spots have been suggested as the main cause for
the required variations in mass transfer rate. 

\begin{figure}
\epsscale{0.9}
\plotone{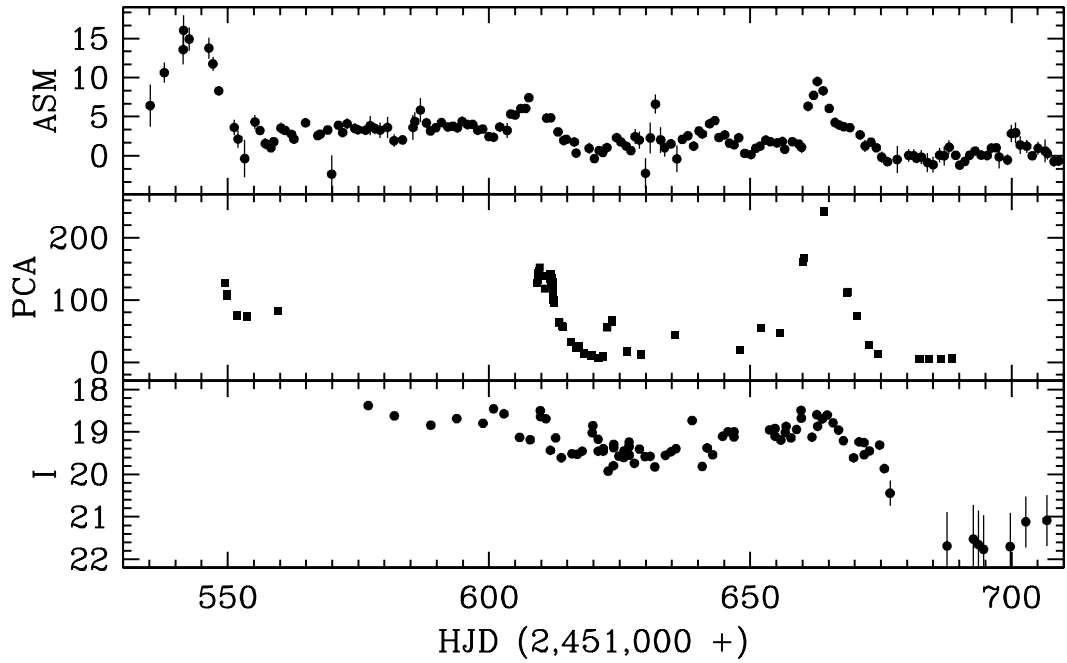}
\epsscale{1.00}
\caption[wachter.f4.eps]{Our 2000 ASM ({\it top}, in counts s$^{-1}$), 
PCA ({\it middle}, in counts s$^{-1}$ PCU$^{-1}$), 
and optical ({\it bottom}) data of X1608$-$52. Note the delay in the optical
upon the return to TQ.  
\label{f-lc2000}}
\end{figure}

In the case of X1608$-$52, it is unclear why the transition to TQ eventually 
occurs and why the 
LIS can last for very different periods of time (between 10 
months and 2 years according to the present RXTE data set). 
Figure~\ref{f-lc2000} shows the details of the transition from the 
LIS to TQ during 2000. The ASM data indicate two plateau-like
levels preceding TQ, each separated by a small X-ray 
``flare'' (HJD 2,451,608 and 
2,451,663). The optical 
data mimic this behavior although the plateaus and the subsequent 
flares are less pronounced, most likely due to the intrinsic 
optical variability during the orbital cycle (see Section 4).
The decay via the observed plateaus may be explained by discrete physical 
changes in the disk as a consequence of a decrease in mass transfer rate. 
The mass 
transfer rate in turn might decrease due to diminishing X-ray irradiation. 
X-ray irradiation is increasingly recognized to play a crucial
role in transient outbursts.
X-ray irradiation stabilizes the accretion disk against the
thermal instability that causes SXT outbursts and increases the
mass transfer rate from the mass donor (Hameury, Lasota, \& Warner 2000; Smak
1995). The LIS might occur because the level of X-ray irradiation in 
X1608$-$52 after 
a major outburst is sufficient to stabilize the mass transfer rate slightly
above the critical level separating persistent and transient behavior.
However, since the system is so close to this critical boundary, any  
change in the X-ray irradiation could trigger a decay to quiescence. 
Such a change in X-ray irradiation might be due to 
the development of transient vertical structure in the disk that shields 
a large fraction of the disk and the mass donor from the central X-ray flux. 
The secondary and disk cool, the
mass transfer rate from the secondary drops, and the disk
temporarily stabilizes at a lower mass transfer level.
However, the new accretion rate and associated X-ray irradiation are now 
insufficient to induce long-term stability in the disk, so the mass transfer declines further and 
the disk eventually reaches the quiescent state. 
The return to TQ in the optical appears to be delayed with respect to the 
X-ray decay. 
Unfortunately, our optical data are very sparse during the transition and 
we can only estimate the delay to be on the order of 2--5 days. 

\section{Periodic Optical Modulation}

\begin{figure}
\epsscale{0.8}
\plotone{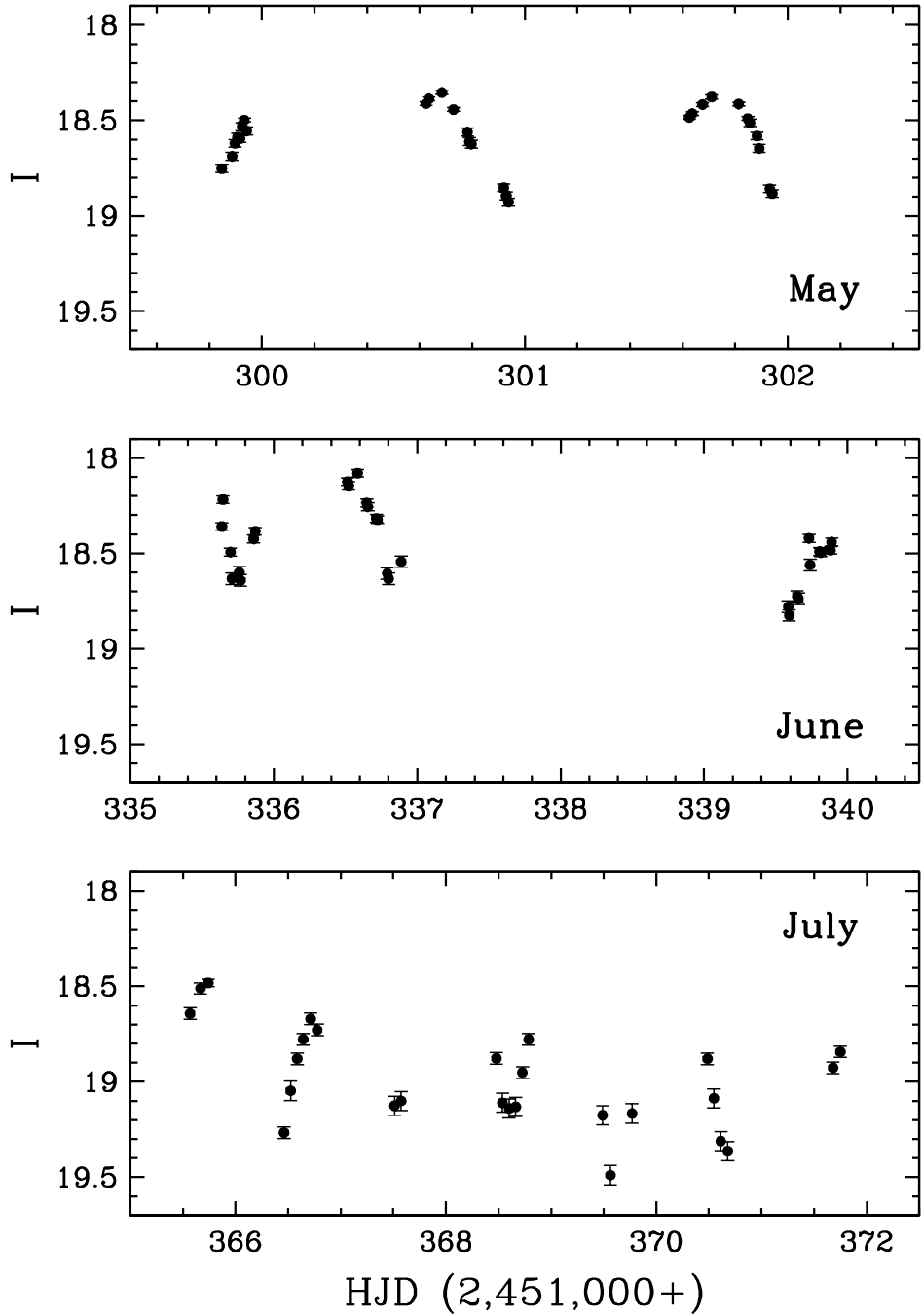}
\epsscale{1.00}
\caption[wachter.f5.eps]{The 1999 $I$ band light curve of QX Nor when 
the system was in the LIS. Large amplitude variations (up to 0.6~mag) 
are visible on 
individual nights as well as changes in the overall mean brightness 
level of the system from night to night.
\label{f-lc1999}}
\end{figure}

Figure~\ref{f-lc1999} shows the optical $I$ band light curve of QX~Nor
obtained in 1999 May, June, and July when the system was in the LIS.  
The source exhibits large amplitude variations of up to 0.6 mag during
each night as well as pronounced changes in overall brightness
between the observations in different months. The May observations
in particular are highly suggestive of a periodic modulation.  
We searched the individual and combined 1999 data sets for a periodic
variability component with CLEAN (Roberts, Lehar \& Dreher, 1987) and 
PDM within IRAF (Stellingwerf 1978). A strong signal for a period 
around 0.5~days is found in the individual data set from each month; 
however, this 
period is not apparent when all the data are combined. More detailed 
analysis shows that this is due to the large changes in the 
overall system brightness. 
Detrending the data (i.e. correcting the data for an overall shift
in the magnitude level for each night) is difficult. 
Simply offsetting by a nightly average does not work due to the large 
amplitude variability and differences in orbital phase coverage on each night.
On nights with 
very few measurements we cannot unambiguously determine which part of 
the light curve is sampled and/or the data do not include reference points 
(either the maximum or minimum transition) that indicate the overall brightness
of the system. We finally assumed that the system was at a 
constant level for all of the May data, and at a different but constant level 
for the first two nights of the June data. These data then define the full 
amplitude between maxima and minima. Subsequently,  
we only corrected those data to the level of the May measurements for which 
either a maximum or minimum transition
is visible and discarded nights with only a few data points for the 
purpose of the period
analysis. For this detrended data set, PDM and CLEAN give best periods
of 0.546$\pm0.059$, 
0.538$\pm0.025$, 0.531$\pm0.022$, 
and 0.5370$\pm0.0015$~days for the May, June, July and combined data 
sets, respectively. 

\begin{figure}
\epsscale{0.8}
\plotone{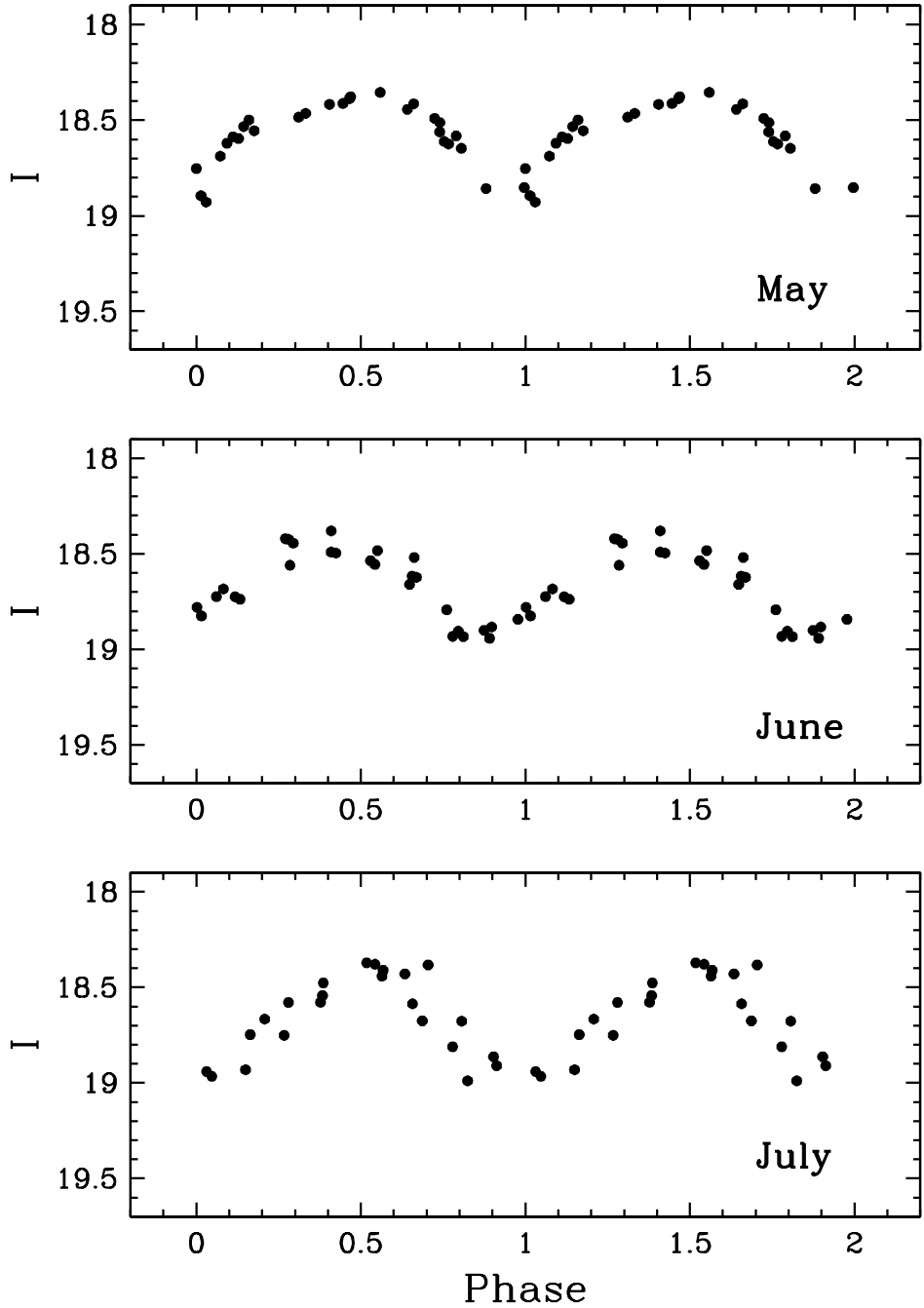}
\epsscale{1.00}
\caption[wachter.f6.eps]{The 1999 optical detrended data set folded on the 
0.5370~day period, with the reference point for the phasing fixed
at the time of the first observation in May. Each data set clearly displays
variability on this period; however, there is also some indication of
a phase shift from one month to the next.
\label{f-fold}}
\end{figure}

Figure~\ref{f-fold} shows the detrended data from each month folded on 
the 0.5370~day period, with the reference point for the phasing fixed 
at the time of the first observation in May. Each data set clearly displays 
variability on this period, however, there is also some indication of 
a phase shift from one month to the next. This phase shift, as well as the 
tendency towards increasingly shorter periods (when each month's data
are considered separately), led us to consider superhumps, in addition 
to more traditional mechanisms, as the cause of 
the photometric variability.  
In the past, photometric modulation 
in LMXBs was thought to be due to either the varying 
aspect of the heated face of the mass donor or 
ellipsoidal variations from the distorted shape of the mass donor.
In both cases, the modulation is strictly periodic and only differs in the 
number of maxima and minima per orbital cycle. Haswell et al.\ (2001) 
discussed possible mechanisms for superhumps in LMXBs and 
claimed that for many LMXBs (with low inclination) the observed 
variability is more likely to be a superhump than an orbital modulation. 
Superhumps are common in CVs, where
they are believed to be the consequence of an eccentric, precessing disk
(Warner 1995). Superhump periods are usually a few percent ($\sim1-7$\%) 
longer than the orbital period and have been found to slowly decrease in length
during outbursts. Aql X-1 may also show superhump behavior. Jain
et al.\ (2001) report stable periodicity in the early stages of
the 1999 outburst of that object at a period of $\approx 18.5$
hours, somewhat smaller than the 18.9 hr period observed by
many authors during outburst, and suggest that the outburst
periodicity may be superhumps.
 
\begin{figure}
\epsscale{0.8}
\plotone{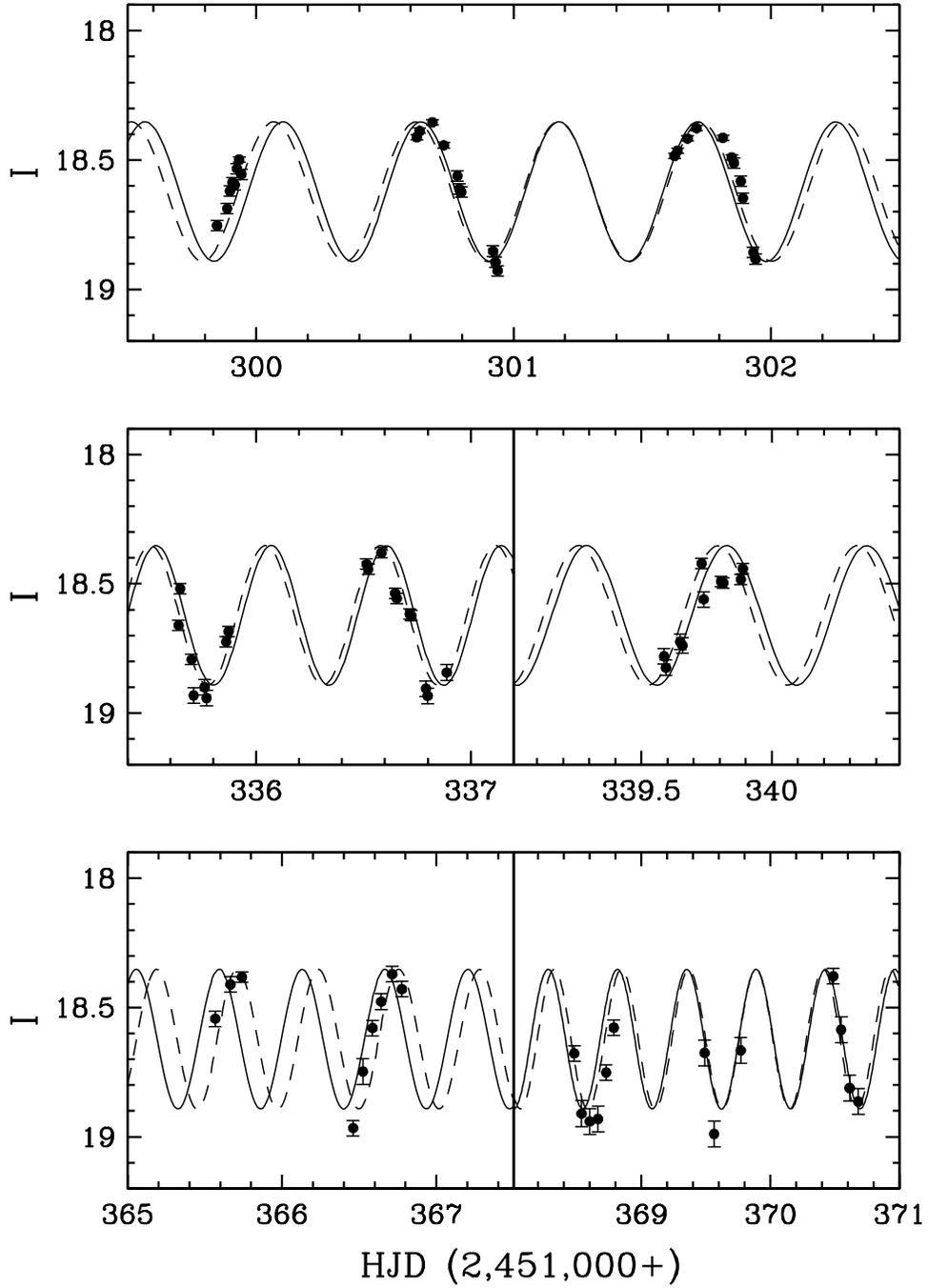}
\epsscale{1.00}
\caption[wachter.f7.eps]{The 1999 detrended $I$ band data set of QX Nor overplotted
with the best fit sine waves for the fixed period (solid line) and the 
time-variable (superhump) period (dashed line) cases.  
\label{f-sh}}
\end{figure}

Figure~\ref{f-sh} shows our detrended light curve data together with two fits,
corresponding to sine waves with either a fixed period of 0.5370~days, 
or with a time variable period of the form $P(t)=P_0 + \dot{P} t$.
In both cases, the sine wave semi-amplitude was held constant at 0.27 mag.
For the variable period, the best fit is achieved with $P_0=0.554$~days and 
$\dot{P}=-0.00024$. The superhump (time-variable) period gives a marginally
better fit with rms $=0.08$ versus rms $=0.12$ for the constant orbital 
period scenario,
but there is clearly not enough data to conclusively distinguish between 
these two possibilities. However, this duplicity could be easily resolved 
during a future bright state of the optical counterpart. 
Several extended monitoring campaigns could be planned in such a way as to 
cover time intervals 
when the two mechanisms produce modulation that is clearly out of phase. 
Alternatively, radial velocity studies in TQ could unambiguously determine
the orbital period of the system, but would require much larger telescopes.
In any case, the orbital period of the system is likely to be close to 
0.537 d, since orbital and superhump periods generally differ by only 
a few percent.  
If the optical emission in the system is dominated by reprocessing of 
X-rays in the disk and the facing side of the mass donor, the detection
of superhump modulation will depend on the inclination of the system according
to Haswell et al.\ (2001). Superhump modulation is expected to
be visible only in low inclination systems, while brightness variations
due to the X-ray heated face of the secondary will dominate in high
inclination systems.
Identifying the process underlying the photometric variability could, 
therefore, offer 
the first clues to the inclination of X1608$-$52.

\section{The Quiescent Counterpart}

\begin{figure}
\epsscale{0.9}
\plotone{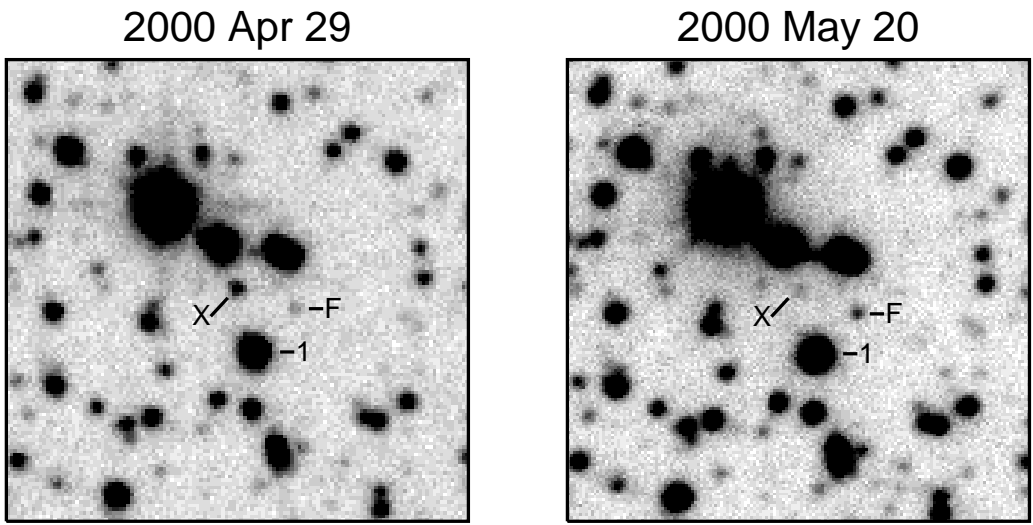}
\epsscale{1.00}
\caption[wachter.f8.eps]{38\arcsec$\times$38\arcsec\ images  of the X1608$-$52 field
obtained with the YALO telescope ({\it left}) when the source was in 
the LIS, and with the CTIO 1.5m telescope ({\it right}) in TQ. `X' marks the 
LMXB, `F' and `1' two local comparison stars.  
Our data set constitutes the first detection of the optical counterpart
in quiescence. 
\label{f-fc}}
\end{figure}

According to the 2000 ASM, PCA, and optical light curves 
(see Figure~\ref{f-lc2000}), X1608$-$52
entered its TQ state around HJD 2,451,680 (2000 May 16).
In the TQ state, photometry of the optical counterpart 
with the YALO (1m) telescope is only possible under exceptional 
observing conditions and with limited accuracy (see Figure~\ref{f-fc}).
Averaging all YALO detections after HJD 2,451,680, we measure $I=21.5\pm0.8$. 
In order to derive more accurate quiescent magnitudes, we also 
observed the counterpart with the CTIO 1.5m telescope on 2000 May 20--22, 
and again in 2001 June 01 with the 
CTIO 4m telescope. The observations on the CTIO 1.5m were obtained 
only a few days after the source faded to TQ. The observations on the 
CTIO 4m telescope, taken over a year later, provide the opportunity 
to determine whether the brightness level detected with the YALO 
after  HJD 2,451,680 indeed represents a stable quiescent state or 
whether the counterpart has faded to even fainter magnitudes 
in the intervening months. 

Figure~\ref{f-fc} shows two $I$ frames of the X1608$-$52 field. 
The image on the left was obtained with the YALO telescope when 
the source was still in the LIS, the image on the right 
was created by combining four 900 s exposures from the CTIO 1.5m in 
the TQ state. Notice the change in brightness relative to the faint 
comparison star F ($I=19.90\pm0.1$). The quiescent counterpart 
is clearly detected in the CTIO 1.5m data with an 
average brightness of $I=21.1\pm0.3$, consistent with the YALO data within 
the errors. Unfortunately, accurate photometry of the faint counterpart 
is very difficult 
since it is strongly affected by the PSF of the nearby bright stars even under 
good seeing conditions. The observations with the CTIO 4m were mainly 
obtained in the $R$ band, since $I$ band observations with the 
MOSAIC II instrument are susceptible
to fringing that is difficult to correct in the crowded 
field of X1608$-$52. We measure $R=22.3\pm0.4$ and $I=20.7\pm0.4$,
confirming that our data actually reflect a stable
quiescent state. 

Previously, Wachter (1997) reported the detection of the IR quiescent 
counterpart and derived constraints on the spectral type of the mass donor. 
We can combine our $R$ and $I$ band measurements with those $J$ and $K$ 
values, and repeat the calculations. We use a distance estimate of 3.6 kpc
based on type I X-ray bursts with photospheric radius expansion
(Nakamura et al.\ 1989). The extinction towards X1608$-$52 is highly 
uncertain. The hydrogen column density ($N_H$) derived from X-ray spectral 
fits in the literature ranges from (1.0--2.0)$\times 10^{22}$ cm$^{-2}$. 
Rutledge et al.\ (1999) use two representative values of 
$N_H = 0.8\times 10^{22}$ cm$^{-2}$ and 
$N_H = 1.5\times 10^{22}$ cm$^{-2}$ for their recent reanalysis of X1608$-$52
X-ray spectra. Utilizing the relationships between $A_V$ and $N_H$ by 
Gorenstein (1975) and Predehl \& Schmitt (1995) leads to 
$4 < A_V < 11$ for the range of $N_H$. A lower limit on the 
extinction can be estimated from the 
observations of the open cluster Ruprecht 115 
by Piatti, Claria \& Bica (1999) which is located only $\sim 3$' away from 
X1608$-$52. The cluster is found to lie at a distance of 2.5$\pm0.7$ kpc with
an extinction of E(B--V)=0.6$\pm0.10$, corresponding to $A_V\approx 2.0$. 
It is also unclear whether the absorbing material is distributed between 
the observer and X1608$-$52 or within the LMXB itself. 

\begin{figure}
\epsscale{0.9}
\plotone{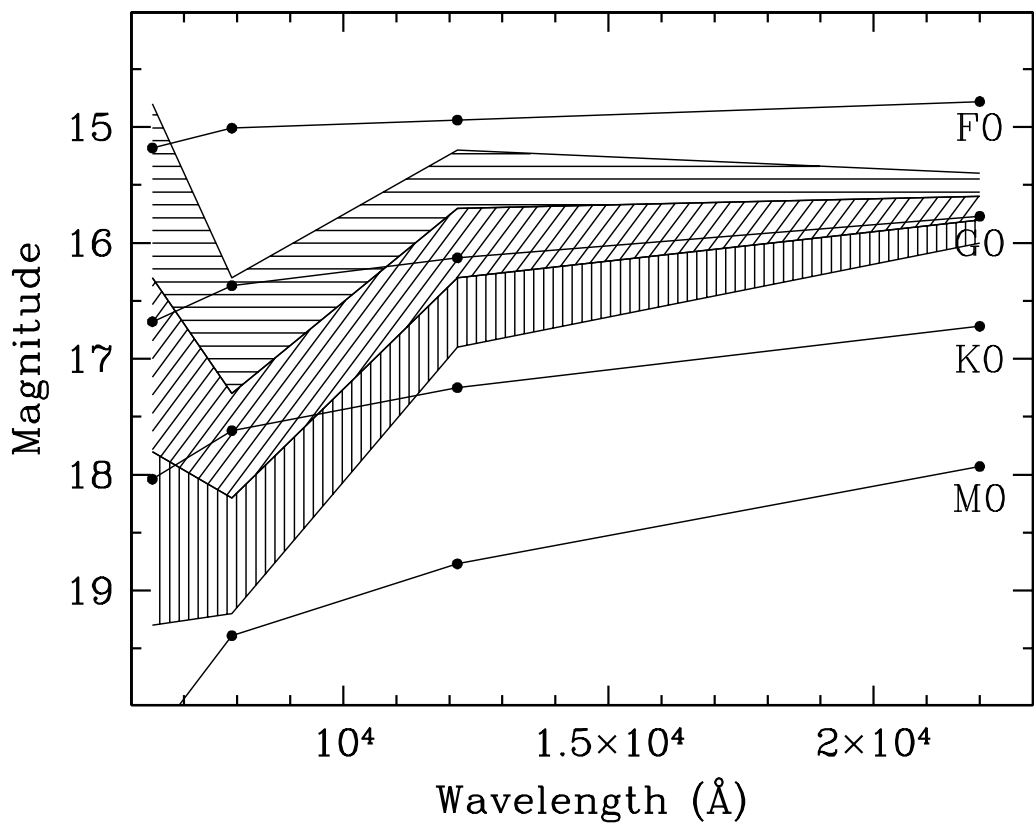}
\epsscale{1.00}
\caption[wachter.f9.eps]{Observational constraints on the spectral
type of the mass donor in X1608$-$52.
The shaded areas indicate the range in dereddened magnitudes for QX~Nor
in the intervals of $A_V=4-6$ (vertical shading),
$A_V=6-8$ (diagonal shading), and $A_V=8-10$ (horizontal shading).
Overplotted are the magnitudes of main sequence stars of four different
spectral types at a distance of 3.6 kpc.
A late F to early G type main sequence star appears to be the most 
consistent with the current data. For an evolved mass donor, the indicated 
spectral types would shift to brighter magnitudes, allowing K and M type secondaries.
\label{f-donor}}
\end{figure}

Figure~\ref{f-donor} illustrates the various photometric constraints 
on the spectral type of the mass donor. The extinction values for the 
different filters were calculated from $A_V$ using the relationships given 
in Cardelli, Clayton, \& Mathis (1989).  
The shaded areas indicate the range in dereddened magnitudes for QX~Nor 
in (arbitrarily chosen) intervals as follows: vertical shading -- $A_V=4-6$ 
($A_R=3-4.5$); diagonal shading -- $A_V=6-8$ ($A_R=4.5-6$); horizontal 
shading -- $A_V=8-10$ ($A_R=6-7.5$). 
We will first consider the case of a main sequence donor. 
The spectral energy distributions of
four different main sequence spectral types at a distance of 3.6 kpc 
are shown in Figure~\ref{f-donor}. 
A late F to early G type main sequence star appears to be the most 
consistent with the current data. One problem is the apparent lack of a 
unique extinction value associated with this choice. However,  
it is important to keep in mind that all the quiescent flux measurements
of QX~Nor carry a rather large uncertainty. In addition, 
a residual accretion disk could contribute some flux and thus
distort the quiescent spectrum attributed to the donor star. 
As seen in the previous section, 
the source also varies strongly during the LIS. It is not 
clear what level of variability can be expected in TQ, 
but it adds another source of uncertainty to the quiescent measurements, which 
were taken years apart in some cases.
The IR magnitudes place the strongest constraints
on the spectral type. It would be important to repeat those measurements,
as well as those at the other wavelengths, with higher accuracy and closer
proximity in time.  

While a giant mass donor is inconsistent with the quiescent magnitudes 
of X1608$-$52, we 
have to consider the possibility that the mass donor is 
evolved. In many LMXBs, both persistent and transient, main sequence donors would
significantly underfill their Roche lobes. Consequently, somewhat evolved 
companions are required. For example, Shahbaz, Naylor \& Charles (1993) conclude
that their observations of the neutron star SXT Cen X-4 are most consistent
with a subgiant mass donor. 
Steeghs \& Casares (2001) show that the mass donor in Sco X-1 must be 
a significantly evolved subgiant.  
In addition, theory suggests that most NS systems require evolved mass 
donors in order to appear as transient systems (King, Kolb \& Burderi 1996).
In Figure~\ref{f-donor}, the spectral energy distributions for the various 
spectral types would shift to brighter magnitudes for evolved secondaries 
(we neglect for the moment any changes in colors due to the uncertainty of 
colors for subgiants), thus pushing a lower mass K star into the regime of 
possible mass donors. 
Using equation (10) from King, Kolb \& Burderi (1996), we find
$M_2/M_2(MS) < 0.2$ as the condition for X1608$-$52 to exhibit SXT behavior
(assuming $M_1$=1.4$M_\odot$ and P=0.537 d). $M_2/M_2(MS)$ is the ratio of 
the secondary mass $M_2$ to that of a main sequence star $M_2(MS)$ of equal 
density which would fill its Roche lobe in a binary of the same period. The 
mean density in a Roche-lobe filling star is determined solely by the 
binary period, $\rho \cong 115 P_{hr}^{-2}$ g cm$^{-3}$ (Frank, King, 
\& Raine 1992). Assuming P=0.537 d, then $\rho = 0.69$ g cm$^{-3}$, which would 
correspond to an F0 main sequence star with $M_2(MS)=1.6 M_\odot$ (Cox 2000).
Hence, we would derive $M_2 = 0.32 M_\odot$ for the mass of an evolved 
mass donor
in X1608$-$52 that fulfills the condition for SXT behavior 
as given by King, Kolb \& Burderi (1996). 

\begin{figure}
\epsscale{0.9}
\plotone{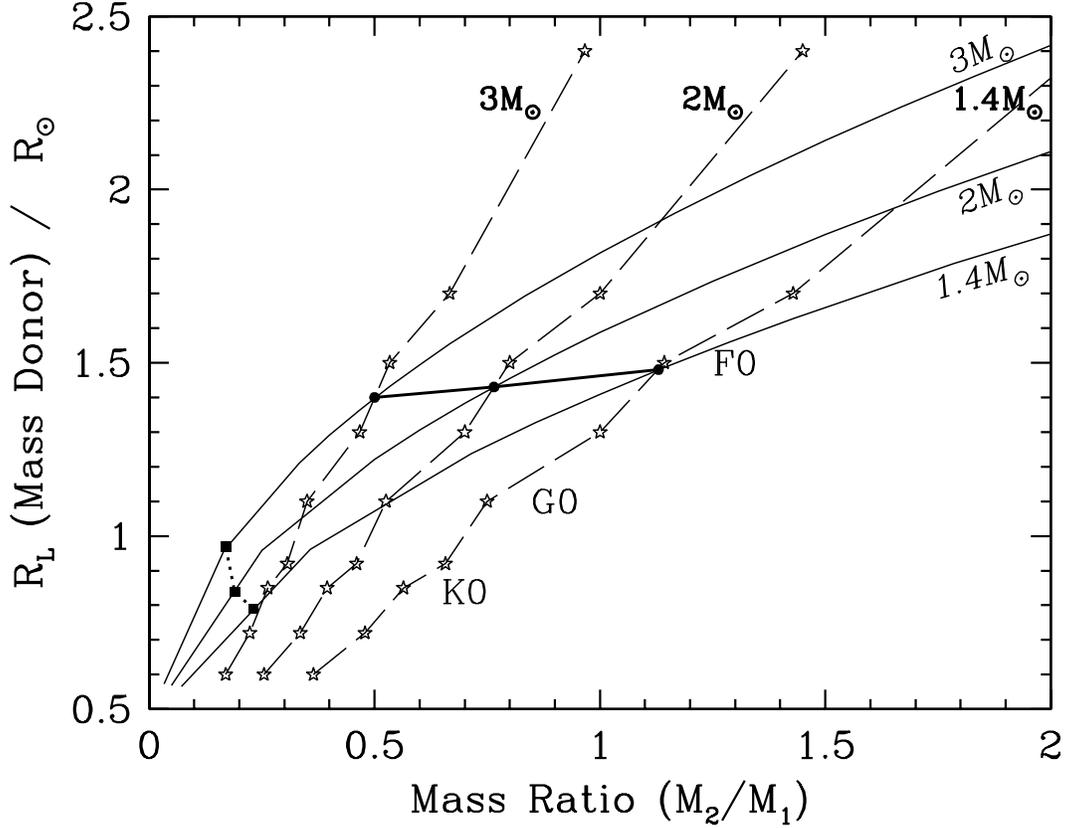}
\epsscale{1.00}
\caption[wachter.f10.eps]{The relationship
between the size of the mass donor's Roche lobe and the mass ratio for
three different NS masses (narrow solid lines). The radii of main sequence
stars for various spectral types (open stars) as a function of mass ratio 
are also 
shown (dashed lines). Each curve is labeled with the mass of the primary (M$_1$) 
used to calculate the mass ratio. The intersection between the two curves 
for fixed values of M$_1$ marks the main sequence spectral type which fills
its Roche lobe in a $\sim0.5$~day binary (solid points, thick solid line).
The data for an evolved mass donor are indicated in the lower left corner of 
the diagram (solid squares, dotted line).
\label{f-rl}}
\end{figure}

Figure~\ref{f-rl} illustrates the relationship
between the size of the mass donor's Roche lobe and the system's mass ratio 
assuming
three different NS masses.
We calculated the Roche lobe size in a 0.5~day binary for a range of 
mass ratios $q$ ($=M_2/M_1$) using the standard approximate analytic
formula by Eggleton (1983),
\begin{equation}
\frac{R_L}{a} = \frac{0.49 q^{2/3}}{0.6 q^{2/3} + ln(1+q^{1/3})}
\end{equation} 
together with Kepler's law. For comparison, the    
radii of main sequence 
stars of various spectral types (taken from Cox 2000) are also shown. 
As calculated above, a main sequence F0 star would fill its Roche lobe in a 0.5~day orbit
with a 1.4$M_{\odot}$ NS. Slightly
later F types are sufficient for higher mass NSs. This is a somewhat
earlier spectral type than favored by the quiescent magnitudes. However,
Figure~\ref{f-donor} shows that a higher value for $A_V$ and/or 
a slightly larger distance would improve the agreement. 
The 3.6 kpc distance to the source was derived assuming specific values for the
mass and radius of the NS and the anisotropy factor of burst emission
(Nakamura et al.\ 1989) and allows variations of 20--40\%.
We have also indicated the case for an evolved mass donor, by repeating the 
calculations detailed in the previous paragraph for different masses of the NS. 
We note that only the case of an evolved mass donor will 
satisfy the condition $q \lesssim 0.33$ required for the formation 
of superhumps if the disk eccentricity is caused by the 3:1 orbital 
resonance.    

\section{Conclusions}

We have identified three distinct states in the X-ray and optical light curves 
of X1608$-$52. In addition to the previously recognized outburst and (true)
quiescence states, X1608$-$52 also exhibits an extended low intensity 
state during which the optical counterpart, QX~Nor, is found to be 
about two magnitudes brighter than during true quiescence. The LIS is 
characterized by the presence of a luminous accretion disk in the system.
The optical counterpart displays pronounced variability that can be 
attributed to either the varying aspect of the X-ray irradiated face of 
the mass donor with an orbital period of 0.5370~days, or a time variable 
superhump period with $P_0=0.554$~days and $\dot{P}=-0.00024$. This ambiguity
could be resolved with photometric monitoring during a future bright 
state of the system. If the mass donor is assumed to be a main sequence star,  
an F to G type star is consistent with the current observations. However,
theoretical considerations argue for an evolved mass donor with a mass
of $\sim 0.32 M_\odot$ (for a $1.4 M_\odot$ NS primary). Only an evolved
mass donor would satisfy the condition for the occurrence of superhumps. 

\acknowledgments

We thank Dr. Alan Levine from the MIT ASM team for help with questions 
regarding the ASM data.
This research has made use of the Simbad
database, operated at CDS, Strasbourg, France. 
Some of the data were obtained as part of the Synoptic, Service, and 
Target-of-Opportunity
(SSTO) program provided by the National Optical Astronomy Observatory 
on the Cerro
Tololo Inter-American Observatory 0.9-m telescope. 
CDB acknowledges support from NSF grant AST 97-30774.

\clearpage

\clearpage

\begin{deluxetable}{cccc}
\tablecaption{Log of observations. \label{tab1}}
\tablewidth{0pt}
\tablehead{
\colhead{Date-Obs (UT)} &
\colhead{Telescope} &
\colhead{Obs} &
\colhead{Filters}
}
\startdata
1999 May 01$-$03 & CTIO 1.5m & 30 & $R$, $I$ \\
1999 Jun 06$-$17 & CTIO 0.9m & 37 & $V$, $I$\\
1999 Jul 06$-$12 & CTIO 0.9m & 26 & $I$ \\
2000 Feb 02$-$Aug 01& CTIO 0.9m\tablenotemark{a} & 55 & $V$, $R$, $I$ \\
2000 Mar 04$-$Jun 11& YALO & 95 & $V$, $I$\\
2000 May 20$-$22    & CTIO 1.5m & 18 & $I$ \\
2001 Feb 07$-$Jun 01& YALO & 24 & $I$ \\
2001 Jun 01  & CTIO 4m & 6 & $R$, $I$ \\
\enddata

\tablenotetext{a}{Some data collected under the SSTO program - see text.}
\end{deluxetable}

\end{document}